\documentclass[12pt,a4]{article}
\usepackage[T1]{fontenc}
\usepackage[utf8]{inputenc}
\usepackage{amsmath}

\usepackage[authoryear, round]{natbib}
\usepackage{graphicx}

\usepackage{datetime}

\begin{document}

\title{Moment-based Bayesian Poisson Mixtures for inferring unobserved units}

\author{Danilo Alunni Fegatelli and Luca Tardella}

\date{}

\maketitle

\abstract{We exploit a suitable moment-based characterization of the mixture of Poisson distribution for developing Bayesian inference for the unknown size of a finite population whose units are subject to multiple occurrences during an enumeration sampling stage. This is a particularly challenging setting for which many other attempts have been made for inferring the unknown characteristics of the population. Here we put particular emphasis on the construction of a default prior elicitation of the characteristics of the mixing distribution. We assess the comparative performance of our approach in real data applications and in a simulation study.}

\newpage

\section{Introduction}

We consider the problem of inferring the total number of units in a finite population in the presence of 
count data where during an experiment or an observation stage all the units are 
potentially observable multiple times but only those who are observed at least once are in fact enumerated in the sample. 
This setting is of interest in wildlife conservation when one is willing to infer on the number of yet unobserved animals living in an area 
using the information coming from the repeated detection of the observed units. 
The same setting occurs in many other fields such as in social sciences where the actual size elusive populations needs to 
be properly assessed \citep{bohn:heij:2009},
in software reliability \citep{Lloyd:1999}, in genomics \citep{wang:lind:cui:wall:zhang:depa:2005}, 
biology  \citep{guin:sepu:paul:mull:2014} 
and linguistics \citep{efro:this:1976}.
In ecology the same type of problem, known as species richness problem \citep{bung:fitz:1993, Chao:Bung:esti:2002, Wang:Lind:pena:2005, chao:chun:2014} has received a lot of attention and many alternative models and methods have been proposed.

Let us fix our model setup. Let $N$ denote the finite size of  the population of interest. Indeed 
it is important to clarify from the outset that in the species sampling terminology $N$ is the number of distinct species and 
not the size of the animal/organism population under investigation.
In order to avoid restrictive homogeneity assumptions 
we can assume that all units act independently from each others conditionally on all the individual detection rates 
so that the joint count probability 
can be expressed as follows
$$
p (\mathbf{c} | \mathbf{\lambda})=\prod_{i=1}^{N}\frac{e^{-\lambda_i}\lambda_i^{c_i}}{c_i!} 
$$
where $\mathbf{\lambda}=(\lambda_1 ,\dots ,\lambda_N)$, $\mathbf{c}=(c_1,\dots ,c_N)$.
However, the individual rate parameters can be thought of 
as unobserved heterogeneous latent intensities
assumed to be drawn from a common distribution 
$Q$. This yields a more flexible hierarchical Poisson mixture distribution for which the 
probability of observing a single count equal to $k$ is 
\begin{align} \label{eq2}
p(C_i=k|Q)  =  h(k,Q)  = \int_{0}^{\infty}\frac{e^{-\lambda}\lambda ^{k}}{k!}dQ(\lambda).
\end{align}
Hence, by exchangeability, the joint probability of observing all the counts
of the population 
$c_{i}$, $i=1,\dots ,N$ 
is summarized by the joint probability of the 
sufficient statistics, called \textit{frequency of frequencies},
$$
\mathbf{f} = (f_0, f_1, ... , f_{k}, ..., f_M)
$$ 
where $f_{k}=\sum_{i=1}^{N} I(c_{i}=k)$ represents the number of units whose count corresponds to $k$ 
and $M=\max(c_{i})$. 
Notice that the number $f_{0}$ of units with count equal to zero is 
not available to the observer and is in fact in one-to-one relation with $N$ 
given  $f_1 ,\dots ,f_M$ since
$$
f_0 = N-\sum_{k=1}^{M} f_k = N-n.
$$
Hence, estimating the main parameter of interest  $N$  
is equivalent to  estimating the number $f_0$  
of unobserved units. 
In this hierarchical formulation the likelihood function can be written as follows 
\begin{align}\label{eq1}
L(N,Q;\mathbf{f})\propto \binom{N}{n}\prod_{k=0}^{M}\left[ h(k,Q)\right] ^{f_{k}} 
\end{align}
where $\mathbf{f}=(f_{0},\dots ,f_{M})$ and $Q$ is the mixing distribution for $\lambda$.
In the literature alternative mixtures of Poisson distributions with 
different finite \citep{pledger:2003} or continuous \citep{bohn:diet:kunh:scho:2005} 
parametric mixing distribution have been considered
as well as other nonparametric likelihood-based estimates \cite{norr:poll:1998,Wang:Lind:pena:2005}. 
In 2010, \cite{wang:2010} proposed to consider a Poisson compound gamma model 
estimating the mixture by a nonparametric penalized maximum likelihood approach 
using a least-squares cross-validation procedure 
for the choice of the common shape parameter.  
Other approaches which are worth mentioning are the Abundance-based Coverage Estimator (ACE), lower bounds and their variants \citep{chao:lee:1992, Mao:jasa:2006}. 
From the Bayesian perspective relevant recent references for 
the parametric approach are \cite{Barg:Bung:obje:2010} and \cite{guin:sepu:paul:mull:2014}
from the nonparametric perspective. A rather different sampling perspective stemming from the species sampling sequential 
approach has been put forward in \cite{lijo:mena:prun:2007} and, more recently, in \cite{zhou:fava:walk:2017}. Notice however that in 
\cite{lijo:mena:prun:2007} the size of the population is indeed assumed to be infinite. 
Differently from \cite{guin:sepu:paul:mull:2014} where a nonparametric Dirichlet process prior is 
used for the nuisance $Q$ our proposal yields an alternative nonparametric estimate of the 
population size based on the likelihood in \eqref{eq1} reparameterized 
in terms of a finite number of moments of a suitable mixing distribution as illustrated in the next section.

\section{Moment-based mixtures of truncated Poisson counts}
\label{sec3-moment-based}
To begin with 
we show that, 
in order simplify our task,
\eqref{eq1} can 
be approximated arbitrarily well by a model in which 
the mixing distribution $Q$ has a compact support in $[0,u]$
for a suitable choice of $u$. 
In fact, the following holds: \\

\textbf{Theorem}: Let $Q$ be a generic probability distribution with support on $[0,\infty)$; $\forall \: \eta >0 \: \exists \: u_{\eta,Q}>0$ such that
\begin{align*}
d_{TV}\left( h(\cdot\: ,\: Q),h(\cdot\: ,\:Q_{u_{\eta,Q}})\right)\leq \eta
\end{align*}
where $Q_{u_{\eta,Q}}$ is the distribution $Q$ restricted to have compact support on $[0,u_{\eta,Q}]$
\\
\textit{proof}: In order to prove the theorem we have to verify that 
$$
\forall \:\eta >0 \:\exists \:u_{\eta,Q} \: :
|Q(A)-Q_{u_{\eta,Q}}(A)|\leq \eta \quad \forall A\in \mathcal{B}(\mathcal{R}^+)
$$
where $\mathcal{B}$ is the Borel $\sigma$-algebra.
Since 
\begin{align}
Q_{u_{\eta,Q}}(A) = \frac{Q\left( A\cap [0,u_{\eta,Q} ] \right)}{Q\left( [0,u_{\eta,Q} ] \right) } \geq Q\left( A\cap [0,u_{\eta,Q} ] \right)
\label{th.1}
\end{align}
and 
\begin{align}
\forall \:\varepsilon(\eta)=\frac{\eta}{1+\eta} >0 \: ; \: \exists \:u_{\eta,Q} : Q([0,u_{\eta,Q} ]) > 1-\varepsilon(\eta) \:\:\Rightarrow \:\: Q([0,u_{\eta,Q} ]^c)<\varepsilon(\eta)
\label{th.2}
\end{align}
we have
\begin{align*}
& Q(A)-Q_{u_{\eta,Q}}(A)= Q\left( A\cap [0,u_{\eta,Q} ] \right) + Q\left( A\cap [0,u_{\eta,Q} ]^c \right) - Q_{u_{\eta,Q}}(A) \leq \\
& Q\left( A\cap [0,u_{\eta,Q} ] \right) + Q\left( [0,u_{\eta,Q} ]^c \right)- Q\left( A\cap [0,u_{\eta,Q} ] \right)<\varepsilon(\eta)<\eta
\end{align*}
Moreover, from \eqref{th.1} and \eqref{th.2} it follows that
\begin{align*}
& Q_{u_{\eta,Q}}(A)-Q(A)= \frac{Q\left( A\cap [0,u_{\eta,Q} ] \right)}{Q\left( [0,u_{\eta,Q} ] \right)}- (Q\left( A\cap [0,u_{\eta,Q} ] \right) + Q\left( A\cap [0,u_{\eta,Q} ]^c \right)) \leq \\
&\frac{Q\left( A\cap [0,u_{\eta,Q} ] \right)}{1-\varepsilon(\eta)}- Q\left( A\cap [0,u_{\eta,Q} ] \right)\leq Q\left( A\cap [0,u_{\eta,Q} ] \right) \frac{\varepsilon(\eta)}{1-\varepsilon(\eta)} \leq \eta
\end{align*}\\
{\phantom{m}} \hfill $\diamondsuit$ \\

This minimal restriction 
on a compact support of the mixing distribution $Q$ 
allows us to 
consider the one-to-one correspondence
of a compact supported univariate distribution $Q_{u}$ and the infinite sequence of its moments.
In fact, we can simplify the  functional form 
of the likelihood as a function of a finite number of characteristics of $Q_{u}$.
To make it explicit
we will be using first another one-to-one mapping 
between finite measures 
\begin{align*}
dQ_{u}(\lambda)=e^{\lambda}dG_{u}(\lambda)
\end{align*}
so that we can eventually regard the likelihood as a function 
of a finite number of moments of the 
finite measure 
$G_u(\cdot)$ uniquely corresponding to $Q_u(\cdot)$. 
Hence,
for a fixed value $u$, we can always consider the 
following simplified parametric model for the probability of each frequency counts 
\begin{align} \label{eq3}
h(k;Q_{u}) =
\int_{0}^{u}\frac{e^{-\lambda}\lambda ^{k}}{k!}dQ_{u}(\lambda)
=\frac{1}{k!}\int_{0}^{u}\lambda^{k}dG_u(\lambda)=\frac{m_{k}(G_{u})}{k!}=h(k;G_{u})
\end{align}
where $m_{k}(G_u)$ is the $k$-th ordinary moment 
corresponding to the finite measure 
$G_u$ not necessarily with total mass equal to 1.
Indeed we can derive the corresponding likelihood
\begin{align}\label{eq4}
L(N,G_u;\mathbf{f})\propto 
\binom{N}{n}\prod_{k=0}^{M}\left[ h(k,Q_u)\right] ^{f_{k}}  
= 
\binom{N}{n}\prod_{k=0}^{M}\left[\frac{m_{k}(G_{u})}{k!} \right] ^{f_{k}}  
\end{align}
which can be thought of 
as an approximate version of the original mixture of Poisson 
model 
\eqref{eq1}.
This suggests that the 
representation of the original model 
in terms of an infinite-dimensional  functional parameter $Q$ 
will be amenable to a 
flexible finite dimensional representation. 
This will ease the task of 
implementing a default Bayesian approach for 
making inference on the parameter of interest $N$. \\
Indeed, in order to further simplify the likelihood structure and represent its
expression as a function of the moments of a 
\textit{probability} measure (with fixed total mass equal to 1)
supported on $[0,u]$ 
we will consider the following trick: 
we take the normalized 
probability distribution  $\tilde{G}_{u}$ 
corresponding to $G_u$, namely 
$$
\tilde{G}_{u} (\cdot) = \frac{G_u(\cdot) }{\int_{0}^{u} dG_u(\lambda) }
$$
so that 
\begin{align*}
\begin{cases}
&m_{0}(\tilde{G}_{u})=\int_{0}^{u} d\tilde{G}_{u}(\lambda)=1 \\
\\
&m_{k}(\tilde{G}_{u})= 
\frac{m_{k}(G_{u})}{m_{0}(G_{u})}
\end{cases}
\end{align*}
It is immediate to realize that since 
$m_{0}(\tilde{G}_{u})=1$
we get 
\begin{align*}
h(k,\tilde{G}_{u})=\frac{1}{k!}\int_{0}^{u}\lambda^{k}d\tilde{G}_{u}(\lambda)=c \cdot h(k,Q_{u}) \quad  \: k=0,\dots ,M
\end{align*}
so that, summing up over all $k$ the normalizing constant 
$c$ is such that
\begin{align*}
c=\sum_{k=0}^{\infty}h(k,\tilde{G}_{u})=\frac{1}{m_{0}(G_{u})}=\frac{1}{h(0,G_{u})}=\frac{1}{\int_{0}^{u} dG_{u}(\lambda)}.
\end{align*}
One can 
replace the use of 
$h(k,Q_{u})$ with 
$c h(k,\tilde{G}_{u})$
and escape from the infinite summation 
defining from the latter expression a 
convenient 
further 
approximation which represents a flexible
parametric distribution for the frequencies of counts 
as follows 
\begin{align}
\label{ourmodel}
h(k,\boldsymbol{m}_{u,M^*})=
\frac{m_{k}(\tilde{G_{u}})}{k!\sum_{j=0}^{M^*}\frac{m_{j}(\tilde{G_{u}})}{j!}} \qquad k=0,\dots , M^*
\end{align}
where the probabilities $h(k,\boldsymbol{m}_{u,M^*})$ are expressed as a function 
of the first $M^*$ 
moments of the probability distribution $\tilde{G}_{u}$
$$
\boldsymbol{m}_{u,M^*}=(m_{u,1},\dots m_{u,k}, \dots ,m_{u,M^*})
$$
where 
$$
m_{u,k}=m_{k}(\tilde{G}_{u})=\int_{0}^{u} \lambda^k d\tilde{G}_{u}(\lambda)
$$
Usually $M^*=M$ but the parametric model is still 
well defined also for $M^* \neq M$.
However, we point out that for the structure of the  likelihood function \eqref{eq1} 
there is information only for the first $M$ moments of the mixing distribution.
The resulting 
model likelihood will be represented as 
\begin{align}\label{eq5}
L(N,\boldsymbol{m}_{u,M^*};\mathbf{f})\propto 
\binom{N}{n}\prod_{k=0}^{M^*}\left[
\frac{m_{u,k}}{k!\sum_{j=0}^{M^*}\frac{m_{u,j}}{j!}}
\right] ^{f_{k}}  
\end{align}
and it can be considered 
a convenient approximation of 
\eqref{eq4} and hence 
of the original 
nonparametric model \eqref{eq1}.
We can make a final simplification
by separating the 
dependence of $m_k(\tilde{G}_u)$
from $u$ 
and the moments
of a single probability distribution
$\tilde{G}_1$
supported on $[0,1]$ namely
\begin{align}\label{ourmodel2}
m_{k}(\tilde{G}_{u})=u^{k}m_{k}(\tilde{G}_{1})
\end{align}
which corresponds to the change of measure for $\tilde{G}_{u}$ due to 
a scale factor $u$ for the rate parameter $\lambda$.
In the following we will use the notation 
$m_k$ instead of $m_{k}(\tilde{G}_{1})$
and $\mathbf{m}_{M^*}=(m_1,...,m_{M^*})$
will be the vector of the first $M^*$ 
moments of an arbitrary probability distribution $\tilde{G}_{1}$ supported on $[0,1]$. 
We can then express our flexible parametric model in terms of a vector of parameters
$(N, \mathbf{m}_{M^*} ,u)\in \{ n,n+1,\dots \}\times {\cal M}_{M^*} \times [0,\infty)$ so that
\begin{align}
\label{eq5}
L(N,\boldsymbol{m}_{M^*},u;\mathbf{f})\propto 
\binom{N}{n}\prod_{k=0}^{M^*}\left[
\frac{u^{k}m_{k}}{k!\sum_{j=0}^{M^*}\frac{u^{j}m_{j}}{j!}}
\right]^{f_{k}}  
\end{align}
where the $M^*$- truncated moment space ${\cal M}_{M^*}$ is such that
$$
{\cal M}_{M^*} =\left\lbrace (m_1 ,\dots , m_{M^*}) : m_k=\int_0^1 x^k d\tilde{G}_1(x) \:,\: \tilde{G}_1\in \mathcal{P}([0,1])\right\rbrace 
$$
where $\mathcal{P}([0,1])$ is the class of probability distributions with support in $[0,1]$.
The ordinary moment space ${\cal M}_{M^*}$ is a constrained $M^*$-dimensional
convex body and hence it is not easy to deal with.
As proposed in \cite{tard:2002} and also used in \cite{tard:farc:08} in the context of 
the discrete-time capture-recapture experiments
one can also consider a further reparameterization of $\boldsymbol{m}_{M^*}$ 
in terms of the so-called canonical moments $\boldsymbol{c}_{M^*}=(c_1 , \dots ,c_{M^*})\in [0,1]^{M^*}$ 
\citep{ski86,det:stu98}.
We define the $k$-truncated moment
class of distributions
$$
\mathcal{P}_{\boldsymbol{m}_{k}}=\left\lbrace 
\tilde{G}_1\in \mathcal{P}([0,1]): \int_{0}^{1}x^{r}d\tilde{G}_1(x)=m_{r}\; , \; r=1,\dots ,k
\right\rbrace 
$$
where $\boldsymbol{m}_{k}=(m_{1},\dots ,m_{k})$.
Moreover, we define the following quantities
\begin{align*}
&m_{k+1}^{+}(\boldsymbol{m}_{k})=\sup_{\tilde{G}_1\in\mathcal{P}_{\boldsymbol{m}_{k}}} m_{r+1}\\
&m_{k+1}^{-}(\boldsymbol{m}_{k})=\inf_{\tilde{G}_1\in\mathcal{P}_{\boldsymbol{m}_{k}}} m_{r+1}
\end{align*}
The generic element $c_{k}$ of $\boldsymbol{c}_{M^*}$ is defined as follows
$$
c_{k}=\frac{m_{k}-m_{k+1}^{-}(\boldsymbol{m}_{k})}{m_{k+1}^{+}(\boldsymbol{m}_{k})-m_{k+1}^{-}(\boldsymbol{m}_{k})} \qquad k=1,\dots ,M^*
$$
so that $\boldsymbol{c}_{M^*}$ can be any point in the space ${\cal C}_{M^*}=[0,1]^{M^*}$.
Then one can do all the computations and simulations 
in this unconstrained parameter space ${\cal C}_{M^*}$ and finally reparameterize back into the space
of the ordinary moments with little extra effort 
so that MCMC approximations of the posterior distribution can be safely derived. 
In order to implement a fully Bayesian approach we need 
to set up a suitable prior distribution for the vector of parameters 
involved in the model.
In the next section we will give details on how one can 
elicit a suitable default prior distribution on the moment space ${\cal M}_{M^*}$.

\section{Default Bayesian inference}
\label{sec3-reference-bayes}

In order to implement a fully Bayesian approach for \eqref{eq5} we need to elicit 
the joint prior distribution for the whole parameter vector ($N, u, m_{1}, \dots ,m_{M^*}$).
We first show how a principled default  Bayesian inference 
can be derived for model \eqref{eq5}  based on 
the count frequency probabilities $h(k,\boldsymbol{m}_{u,M^*})$.\\

We note that, 
for fixed values of the parameters $N$ and $u$ taking 
$n_0=N-\sum_{k=1}^{M^*} n_k$ 
%%%
% I LEFT OUT LO DICIAMO CHE VENGONO POI AGGIUNTI ALLA STIMA DI N? VEDI SCRAPIE EXAMPLE
%%%
the expression in
\eqref{eq5} is a multinomial likelihood 
in terms of the probabilities $\boldsymbol{h}_{M^*} = h(0,\boldsymbol{m}_{u,M^*}), \dots , h(M^*,\boldsymbol{m}_{u,M^*})$ 
which are in turn one-to-one related to $\boldsymbol{m}_{M^*}=(m_{0}, \dots , m_{M^*})$.
This allows us to consider a standard Jeffreys'prior on 
$\boldsymbol{h}_{M^*}$
and transform it back in terms of a default distribution on 
$\boldsymbol{m}_{M^*}$ 
conditionally on any fixed value of $N$ and $u$, taking into account the appropriate Jacobian.
It is known that the Jeffreys'prior for an unconstrained multinomial 
parameter vector is a Dirichlet distribution 
and one can argue that for the count frequency probabilities which are constrained 
on a proper convex body contained in the $M^*$-dimensional simplex the same 
functional form of the Jeffreys' prior is preserved up 
to a different normalizing constant.
So we have
\begin{align}\label{prior-f}
\pi_J (h(1;\boldsymbol{m}_{u,M^*}),\dots , h(1;\boldsymbol{m}_{u,M^*}))\propto \prod_{k=0}^{M^*}\left[ h(k;\boldsymbol{m}_{u,M^*})\right] ^{-\frac{1}{2}}
\end{align}
As previously mentioned simulation within the moment space can be eased by reparameterizing the ordinary moments of the distribution $\tilde{G}_1\in[0,1]$ 
in terms of the corresponding canonical moments \citep{tard:2002}.
The only step needed to re-express our Jeffreys prior in terms of $m_{1},\dots ,m_{M^*}$ 
is the evaluation of the appropriate Jacobian.
Indeed, to simplify formulae,
let us denote with $x_{k}=h(k,\boldsymbol{m}_{u,M^*})$, $y_{k}=\frac{m_{k}(\tilde{G_{u}})}{k!}$, $\mathbf{x}=(x_1 ,\dots ,x_{M^*})$ and $\mathbf{y}=(y_1 ,\dots ,y_{M^*})$. 
The count frequencies in \eqref{ourmodel} can be expressed as a function
of $\mathbf{y}$: 
$$
\mathbf{x}=g(\mathbf{y})
$$
as follows
\begin{align*}
x_{k}=\frac{y_{k}}{\sum_{j=0}^{M^*}y_{j}}=\frac{y_{k}}{D_{\mathbf{y}}} 
\end{align*}
where $D_{\mathbf{y}}=\sum_{j=0}^{M^*}y_{j}$ stands for the denominator. 
Notice that both vectors $\mathbf{x}$ and $\mathbf{y}$ can be completed
when needed by $x_{0}=f(0,\boldsymbol{m}_{u,M^*})$ and $y_{0}=\frac{m_{0}(\tilde{G_{u}})}{0!}$ 
using the known constraints: $\sum_{k=0}^{M^*}x_{k}=1$ and $y_{0}=1$.  
Hence we have that the standard Jeffreys'prior on multinomial cell probabilities $\mathbf{x}$
is 
%\begin{align*}
%\int_{\mathcal{X}}w(\mathbf{x})d(\mathbf{x})=\int_{\mathcal{Y}}w\left( g(\mathbf{y})\right)\cdot \mid \mathbf{J}_{g(\mathbf{y})} \mid d\mathbf{y}
%\end{align*}
\begin{align*}
\pi_{J}(\mathbf{x})\propto \prod_{k=0}^{M^*}x_{k}^{-\frac{1}{2}}
\end{align*}
and the corresponding Jeffreys'prior in terms of $\mathbf{y}=g^{-1}(\mathbf{x})$ 
can be written as 
\begin{align}
\pi_{J}^{\star}(\mathbf{y})=\pi_{J}(g(\mathbf{y})) \cdot \mid J_{g}(\mathbf{y})\mid
\label{ccc}
\end{align}
where $\mathbf{J}_{g}(\mathbf{y})=[j_{i,j}(\mathbf{y})]$ is the Jacobian matrix containing the partial derivatives of $g(\mathbf{y})$. The Jacobian matrix has the extra-diagonal elements
\begin{align*}
j_{i,j}(\mathbf{y})=-\frac{y_{j}}{D_{\mathbf{y}}^2} \qquad & \forall i\: \forall j ; \; i\neq j
\end{align*} 
while the diagonal elements are
\begin{align*}
j_{i,i}(\mathbf{y})=\frac{D_{\mathbf{y}}-y_{i}}{D_{\mathbf{y}}^2} \qquad i=1,\dots ,M^*
\end{align*} 
Now we finally express the Jeffreys'prior in terms of $\mathbf{m}_{M^*}$ using \eqref{ccc}
and the one-to-one mapping \eqref{ourmodel2} which 
maps $\mathbf{y}$ into $\mathbf{m}_{M^*}$
\begin{align*}
y_{k}=\frac{u^k}{k!} m_{k} \Rightarrow \mathbf{y}=h(\mathbf{m}_{M^*})
\end{align*}
and hence we have
\begin{align*}
\pi_{R}(\mathbf{m}_{M^*})=\pi_{J}(g(h(\mathbf{m}_{M^*})))\cdot \mathbf{J}_{g}(h(\mathbf{m}_{M^*}))  \cdot   \mid \mathbf{J}_{h}(\mathbf{m}_{M^*}) \mid
\end{align*}
where $\mid \mathbf{J}_{h}(\mathbf{m}_{M^*}) \mid$ is easily to derived as follows
$$
\mid \mathbf{J}_{h}(\mathbf{m}_{M^*}) \mid = \prod_{k=1}^{M^*} \frac{u^k}{k!}
$$
To complete the prior elicitation for our model we consider for $N$
three different non-informative prior distributions: 
uniform, $1/N$ and Rissanen's prior. 
We will investigate the sensitivity of the posterior analyses and compare its performances 
by simulation study and results of some real data examples. \\
Notice that so far we have assumed a fixed upperbound $u$ for the support of 
the mixing distribution of $\lambda$.
Now we need to endow $u$ with a prior distribution. 
Indeed considering how we jointly rescale all the moments of $\tilde{G}_1$
into the moments of $\tilde{G}_u$
\begin{align*}
&m_1(\tilde{G}_u)=u \:m_1(\tilde{G}_1)\\
&\dots \\
&m_k(\tilde{G}_u)=u^k \:m_k(\tilde{G}_1)\\
&\dots \\
&m_{M^*}(\tilde{G}_u)=u^{M^*} \:m_{M^*}(\tilde{G}_1)
\end{align*}
we use as a reference distribution
\begin{align}
\pi_R(u)\propto u^{-\frac{M^*(M^*+1)}{2}}
\label{prior-u}
\end{align}
In order to avoid an improper distribution and degenerate inference for 
$u\rightarrow 0$ we fix a positive  lowerbound ($u_{LB}=0.5$)
for the support of $u$.

\section{Applications}

\subsection{Application to Simulated data}

In order to evaluate the performance of our proposal we implemented a simulation study 
according to the same setting considered in \cite{wang:2010} as described in Table \ref{tab.sim.wang}.
\begin{table}[!h]
\begin{center}
\begin{tabular}{clc}
  \hline
\quad \textbf{Setting} \quad & \textbf{Distribution} ($Q$) & \quad $E(M/n)$ \quad \\
\hline \\
& \textbf{Gamma} & \\
\quad 1 \quad 	&  $Ga(4, 3.125)$ & 0.90\\
\quad 2 \quad 	&  $Ga(4, 1)$ & 0.59\\
\quad 3 \quad 	&  $Ga(1, 0.25)$ & 0.20 \medskip
\\
& \textbf{Gamma Mixture}\\
\quad 4 \quad 	&  $0.5 \cdot Ga(2, 1) + 0.5 \cdot Ga(2, 2)$ & 0.65\\
\quad 5 \quad 	&  $0.5 \cdot Ga(2, 1) + 0.5 \cdot Ga(4, 1)$ & 0.57 \medskip
\\
& \textbf{Log-Normal} & \\
\quad 6 \quad 	&  $LN(0.75, 0.75)$ & 0.82\\
\quad 7 \quad 	&  $LN(-0.5, 2)$ & 0.50\\
\quad 8 \quad 	&  $LN(-1, 1)$ & 0.36 \medskip
\\
& \textbf{Log-Normal Mixture}\\
\quad 9 \quad 	&  $0.5 \cdot LN(-0.5, 1) + 0.5 \cdot LN(0.5, 1)$ & 0.61\medskip
\\
& \textbf{Finite Mixture}\\
\quad 10 \quad 	&  0.8 $\cdot \delta (1.2) + 0.2 \cdot \delta (6.7)$ & 0.76\\
\quad 11 \quad 	&  0.89 $\cdot \delta (0.5) + 0.11 \cdot \delta (6.7)$ & 0.46\\
\quad 12 \quad 	&  0.8 $\cdot \delta (0.2) + 0.2 \cdot \delta (1.3)$ & 0.29\\
  \hline
\end{tabular}
\end{center}
\caption{\textit{Simulation setting (\cite{wang:2010})}}
\label{tab.sim.wang}
\end{table}
For each setting a different mixing distribution on the Poisson intensity
is fixed and 100 simulated datasets are drawn and used 
to repeat the estimation procedure.
Bias and mean square error of point estimates and coverage of interval estimates are 
approximatively evaluated averaging the results 
obtained with the simulated datasets.
We compare 
our method with the recent non parametric approach based on a penalized likelihood 
proposed in \cite{wang:2010} which highlighted inferential
difficulties of the previously available
approaches and showed a substantial improvement over the latter. 
Wang's procedure is implemented in the {\tt{R}} package {\tt{SPECIES}}
\citep{SPECIES:2011}
where the corresponding function is named {\tt{pcg}(\dots )}. 
The package allows also to compute point and confidence interval estimates 
from alternative nonparametric and semi-parametric methods 
using the first $M^*$ counts observed.
In order to make a sound comparison with Wang's procedure we fixed the number of moments 
of the probability distribution $\tilde{G}_{u}$ considered to be $M^*=10$ 
since in Wang's simulation study only the first 10 counts are considered.
Although we evaluated several prior choices for $N$
we report in Table \ref{sim-results} only the results obtained from the uniform prior $\pi(N)\propto 1$
which leads to the best performances. 
We will denote by $\hat{N}_{BPM}$ %and $\hat{N}_{BPM_A}$ respectively
the resulting estimator.
\begin{table}[!h]
\begin{center}
\begin{tabular}{|ccccc|ccccc|}
\hline \footnotesize{\textbf{Setting}} & \footnotesize{$\hat{N}$} & \footnotesize{$\hat{Me}$} & \footnotesize{$MSE$} & \footnotesize{$\% \: Cov$} & \footnotesize{\textbf{Setting}} & \footnotesize{$\hat{N}$} & \footnotesize{$\hat{Me}$} & \footnotesize{$MSE$} & \footnotesize{$\% \: Cov$}\\
\hline
\hline  \footnotesize{\textbf{1}} & \footnotesize{$\hat{N}_{BPM_{}}$} & \footnotesize{1020} & \footnotesize{27.93} & \footnotesize{100} & \footnotesize{\textbf{2}} & \footnotesize{$\hat{N}_{BPM_{}}$} & \footnotesize{1135} & \footnotesize{160.73} & \footnotesize{99} \\
%   & \footnotesize{$\hat{N}_{BPM_{2}}$} & \footnotesize{1020} & \footnotesize{28.19} & \footnotesize{100} &  & \footnotesize{$\hat{N}_{BPM_{2}}$} & \footnotesize{1143} & \footnotesize{162.07} & \footnotesize{99} \\
   & \footnotesize{$\hat{N}_{PL}$} & \footnotesize{1020} & \footnotesize{28.11} & \footnotesize{97} &  & \footnotesize{$\hat{N}_{PL}$} & \footnotesize{1138} & \footnotesize{161.00} & \footnotesize{99} \\
   & \footnotesize{$\hat{N}_{PCG}$} & \footnotesize{1011} & \footnotesize{28.39} & \footnotesize{95} &  & \footnotesize{$\hat{N}_{PCG}$} & \footnotesize{1014} & \footnotesize{149.47} & \footnotesize{99} \\\hline
\hline  \footnotesize{\textbf{3}} & \footnotesize{$\hat{N}_{BPM_{}}$} & \footnotesize{1070} & \footnotesize{147.85} & \footnotesize{100} & \footnotesize{\textbf{4}} & \footnotesize{$\hat{N}_{BPM_{}}$} & \footnotesize{1009} & \footnotesize{58.08} & \footnotesize{100} \\
 %  & \footnotesize{$\hat{N}_{BPM_{2}}$} & \footnotesize{1066} & \footnotesize{149.44} & \footnotesize{100} &  & \footnotesize{$\hat{N}_{BPM_{2}}$} & \footnotesize{1014} & \footnotesize{59.54} & \footnotesize{100} \\
   & \footnotesize{$\hat{N}_{PL}$} & \footnotesize{1034} & \footnotesize{133.25} & \footnotesize{100} &  & \footnotesize{$\hat{N}_{PL}$} & \footnotesize{1013} & \footnotesize{59.16} & \footnotesize{100} \\
   & \footnotesize{$\hat{N}_{PCG}$} & \footnotesize{924} & \footnotesize{234.71} & \footnotesize{100} &  & \footnotesize{$\hat{N}_{PCG}$} & \footnotesize{991} & \footnotesize{124.47} & \footnotesize{99} \\\hline
\hline  \footnotesize{\textbf{5}} & \footnotesize{$\hat{N}_{BPM_{}}$} & \footnotesize{1041} & \footnotesize{72.02} & \footnotesize{100} & \footnotesize{\textbf{6}} & \footnotesize{$\hat{N}_{BPM_{}}$} & \footnotesize{1004} & \footnotesize{106.64} & \footnotesize{100} \\
%   & \footnotesize{$\hat{N}_{BPM_{2}}$} & \footnotesize{1043} & \footnotesize{72.22} & \footnotesize{100} &  & \footnotesize{$\hat{N}_{BPM_{2}}$} & \footnotesize{1011} & \footnotesize{104.40} & \footnotesize{100} \\
   & \footnotesize{$\hat{N}_{PL}$} & \footnotesize{1040} & \footnotesize{72.42} & \footnotesize{100} &  & \footnotesize{$\hat{N}_{PL}$} & \footnotesize{997} & \footnotesize{102.60} & \footnotesize{100} \\
   & \footnotesize{$\hat{N}_{PCG}$} & \footnotesize{1009} & \footnotesize{160.21} & \footnotesize{96} &  & \footnotesize{$\hat{N}_{PCG}$} & \footnotesize{1041} & \footnotesize{113.63} & \footnotesize{98} \\
   \hline
\hline  \footnotesize{\textbf{7}} & \footnotesize{$\hat{N}_{BPM_{}}$} & \footnotesize{829} & \footnotesize{171.03}  & \footnotesize{83} & \footnotesize{\textbf{8}} & \footnotesize{$\hat{N}_{BPM_{}}$} & \footnotesize{907} & \footnotesize{113.89} & \footnotesize{100} \\
%   & \footnotesize{$\hat{N}_{BPM_{2}}$} & \footnotesize{828} & \footnotesize{169.77} & \footnotesize{81} &  & \footnotesize{$\hat{N}_{BPM_{2}}$} & \footnotesize{904} & \footnotesize{116.96} & \footnotesize{100} \\
   & \footnotesize{$\hat{N}_{PL}$} & \footnotesize{831} & \footnotesize{169.51} & \footnotesize{86} &  & \footnotesize{$\hat{N}_{PL}$} & \footnotesize{912} & \footnotesize{115.77} & \footnotesize{100} \\
   & \footnotesize{$\hat{N}_{PCG}$} & \footnotesize{996} & \footnotesize{198.86} & \footnotesize{97} &  & \footnotesize{$\hat{N}_{PCG}$} & \footnotesize{1016} & \footnotesize{197.61} & \footnotesize{99} \\
   \hline
\hline  \footnotesize{\textbf{9}} & \footnotesize{$\hat{N}_{BPM_{}}$} & \footnotesize{976} & \footnotesize{71.94} & \footnotesize{98} & \footnotesize{\textbf{10}} & \footnotesize{$\hat{N}_{BPM_{}}$} & \footnotesize{1117} & \footnotesize{122.48} & \footnotesize{72} \\
%   & \footnotesize{$\hat{N}_{BPM_{2}}$} & \footnotesize{975} & \footnotesize{73.76} & \footnotesize{97} &  & \footnotesize{$\hat{N}_{BPM_{2}}$} & \footnotesize{1118} & \footnotesize{122.94} & \footnotesize{68} \\
   & \footnotesize{$\hat{N}_{PL}$} & \footnotesize{974} & \footnotesize{71.88} & \footnotesize{97} &  & \footnotesize{$\hat{N}_{PL}$} & \footnotesize{1061} & \footnotesize{78.02} & \footnotesize{88} \\
   & \footnotesize{$\hat{N}_{PCG}$} & \footnotesize{1028} & \footnotesize{163.07} & \footnotesize{100} &  & \footnotesize{$\hat{N}_{PCG}$} & \footnotesize{1038} & \footnotesize{56.93} & \footnotesize{83} \\
   \hline
\hline  \footnotesize{\textbf{11}} & \footnotesize{$\hat{N}_{BPM_{}}$} & \footnotesize{1207} & \footnotesize{281.11} & \footnotesize{91} & \footnotesize{\textbf{12}} & \footnotesize{$\hat{N}_{BPM_{}}$} & \footnotesize{880} & \footnotesize{154.43} & \footnotesize{100} \\
%   & \footnotesize{$\hat{N}_{BPM_{2}}$} & \footnotesize{1203} & \footnotesize{283.45} & \footnotesize{88} &  & \footnotesize{$\hat{N}_{BPM_{2}}$} & \footnotesize{880} & \footnotesize{152.99} & \footnotesize{100} \\
   & \footnotesize{$\hat{N}_{PL}$} & \footnotesize{1192} & \footnotesize{276.01} & \footnotesize{91} &  & \footnotesize{$\hat{N}_{PL}$} & \footnotesize{879} & \footnotesize{153.87} & \footnotesize{100} \\
   & \footnotesize{$\hat{N}_{PCG}$} & \footnotesize{1035} & \footnotesize{177.26} & \footnotesize{87} &  & \footnotesize{$\hat{N}_{PCG}$} & \footnotesize{938} & \footnotesize{169.39} & \footnotesize{93} \\
\hline
\end{tabular}
\end{center}
\caption{\textit{Comparing four different estimators with respect to median bias, mean squared error
and 95\% confidence interval coverage in 12 simulation settings listed in Table \ref{tab.sim.wang}}}
\label{sim-results}
\end{table}

\begin{figure}[!h]
\centering
\includegraphics[scale=0.56]{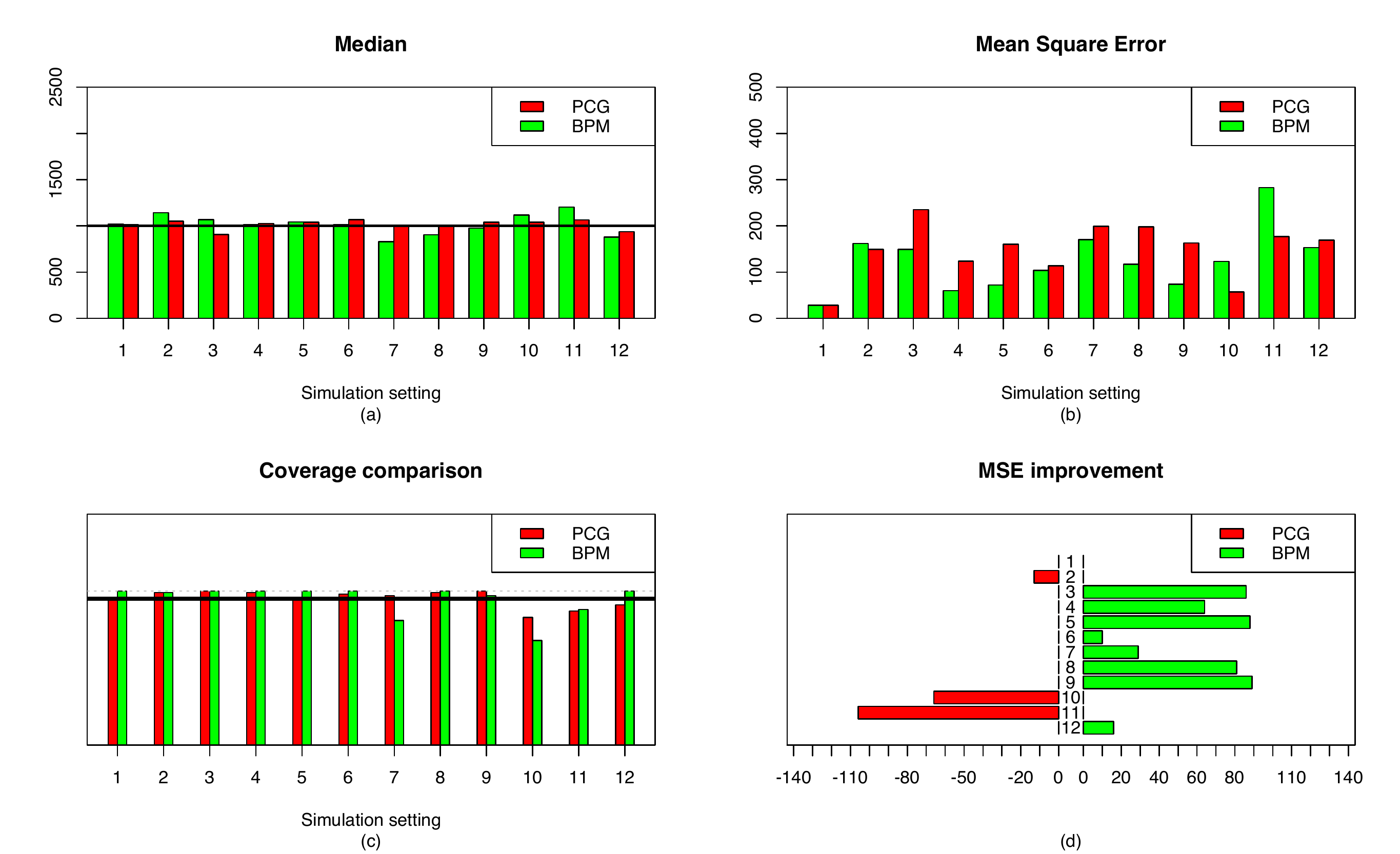}
\caption{\textit{Comparing PCG and fully Bayesian approach: Summary}}
\label{BPM-Wang}
\end{figure}
\begin{figure}[!h]
\centering
\includegraphics[scale=0.25]{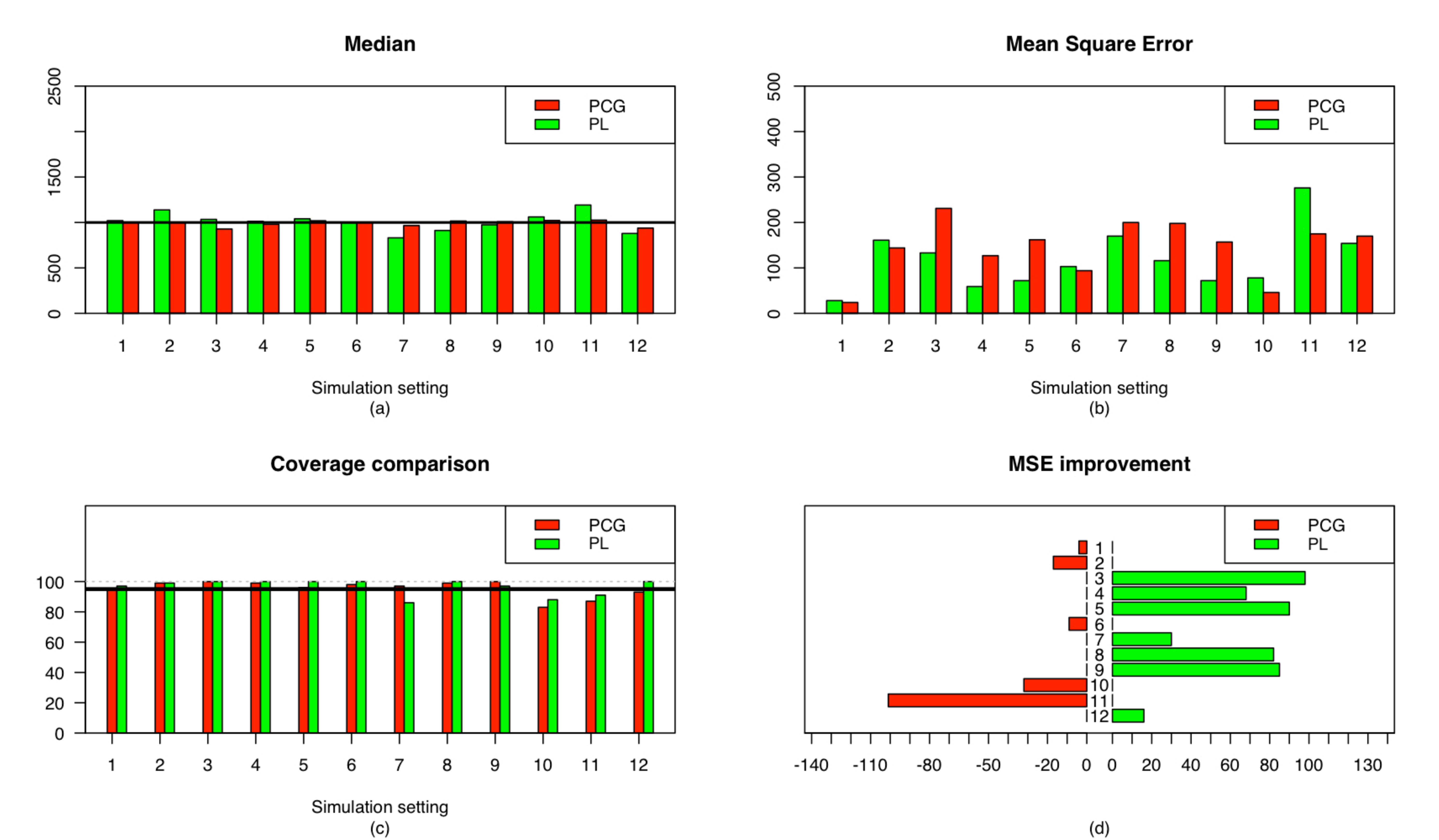}
\caption{\textit{Comparing PCG and integrating a modified/penalized likelihood approach: Summary}}
\label{comp-wang-senza-jac}
\end{figure}

As we can see from the results in 
Table \ref{sim-results} graphically summarized in Figure \ref{BPM-Wang}
our Bayesian estimators seem to compete well with Wang's {\tt{pcg}} procedure although
occasionally they can be beaten in terms of efficiency and interval coverage.
In his paper Wang shows how his estimator almost uniformly outperforms
all previously available estimators in terms of precision and coverage.
We find out that a slight modification of the fully Bayesian recipe can do even better.
It turns out that integrating out the following penalized likelihood 
$$
L_P(N,\boldsymbol{m}_{M^*},u;\mathbf{f})\propto 
\binom{N}{n}\prod_{k=0}^{M^*}\left[
\frac{u^{k}m_{k}}{k!\sum_{j=0}^{M^*}\frac{u^{j}m_{j}}{j!}}
\right]^{f_{k}-\frac{1}{2}} $$
with the
similar prior choices for $N$ and $u$ and a uniform measure on the 
moments $m_1, \dots , m_{M^*}$ one gets a better performance as we can see in  
Figure \ref{comp-wang-senza-jac}.
However, we will not consider it further because it does not correspond to a fully Bayesian approach.\\
Moreover, even though our new methods (fully Bayesian and penalized integrated
likelihood) are computationally intensive, 
the derivation of the interval estimates is often quicker 
compared to Wang's {\tt{pcg}} procedure which relies on a costly double-bootstrap procedure. 
Overall if we average on all the twelve simulation settings 
our $N_{BPM}$ turns out to be an improvement  over $N_{PCG}$ in terms of average mean 
square error while the corresponding interval estimates show
an overall suitable coverage close to the nominal level.

\subsection{Real data analyses}

We investigate the effectiveness of our proposed estimator
with several benchmark datasets used in the recent 
works of \cite{wang:2010} and \cite{rocc:bung:bohn:2011}
comparing our Bayesian approach with both approaches
developed in these papers.
The estimator $\hat{N}_{RBB}$ proposed in
\cite{rocc:bung:bohn:2011} 
is based on a linear regression model
on the ratios of successive frequency
counts. Namely 
$$
\hat{r}(x)=\frac{(x+1)f_{x+1}}{x f_{x}}
$$
We stress that such estimator does not aim to be a flexible 
nonparametric estimator since it is derived under the assumption that
the count distribution belongs to the so called Katz family \citep{katz:1952}.
For this reason we have not used it as alternative competitor in our simulation study.
For the following real data Bayesian analyses we will follow the recipe recommended from 
the simulation study: uniform prior for $N$, 
Jeffreys'prior on $\boldsymbol{m}_{M^*}$ 
and for $u$ we consider the reference prior $\pi_{R}(u)$
described in \eqref{prior-u}.

\subsubsection*{Traffic data}
\begin{figure}[!h]
\centering
%\includegraphics[height=10.7cm, width=25cm, angle=0,
%keepaspectratio]{Comparing-TAF-Wang-2012.pdf}
\includegraphics[scale=0.58]{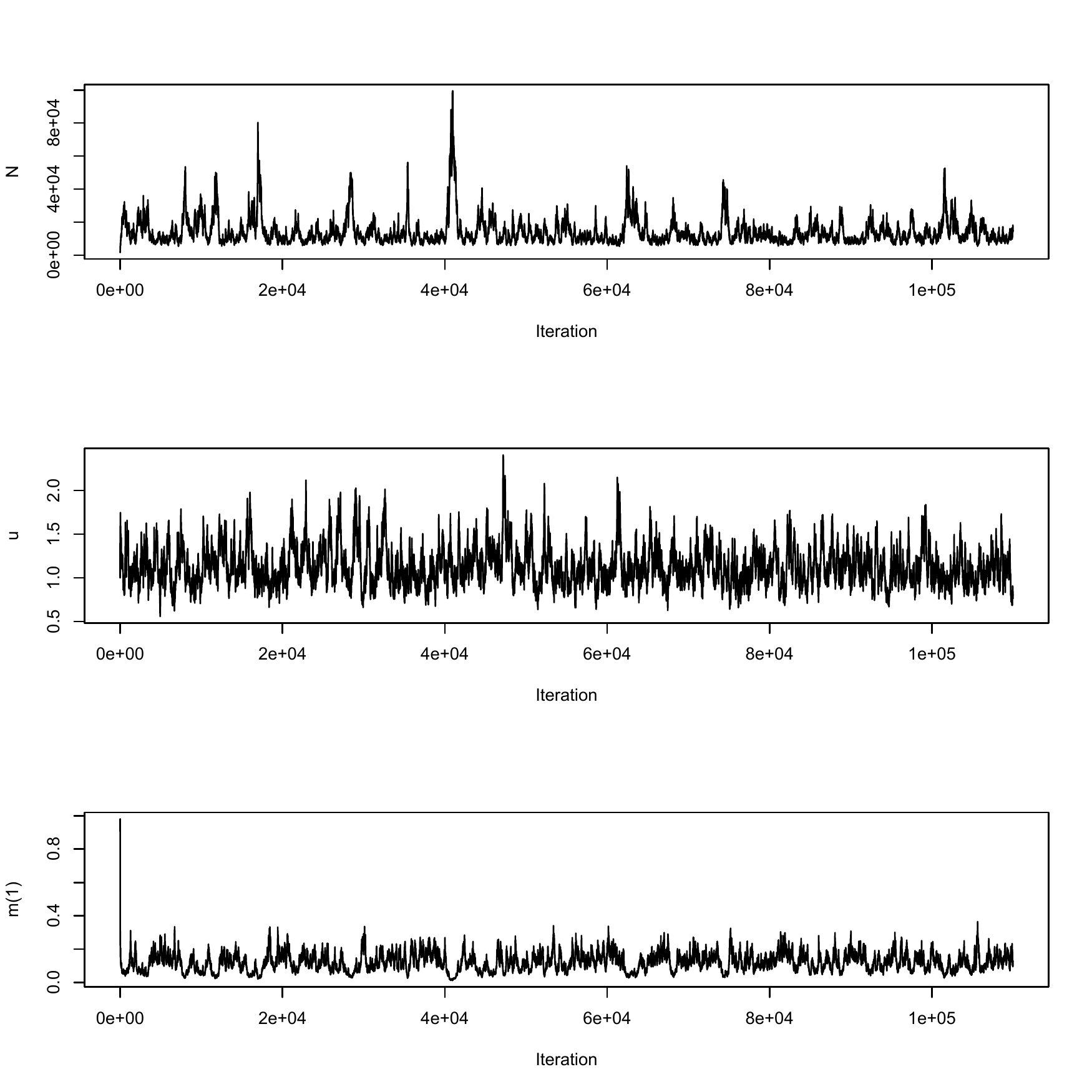}
\caption{\textit{Trace-plot of $N$, $u$ and $m_{1,7}$}.}
\label{trace}
\end{figure}

We start with the famous dataset
known as \textit{Traffic Data} originally 
studied in \cite{sima:1976} and lately re-analyzed in 
\cite{bohn:scho:2005} and \cite{wang:2010}.
%In table ? is shown traffic dataset from Thyrion (1961) and Simar (1976) 
%and also revised in many other papers as B\"{o}hning \& Sch\"{o}n (2005) and Wang (2010). 
Data are shown in Table \ref{traffic-data}. 
They represent the accident counts submitted to 
La Royale Belge Insurance Company during a particular year. 
In this example we know the real value for $N$ ($9461$) which is the 
total number of insurance policies covering both ``business'' and ``tourist'' automobiles; 
hence the complete frequency counts show that 
the proportion of the unobserved units is very high. 
\begin{table}[!h]
\begin{center}
\begin{tabular}{|ccccccccc|}
\hline
$k$ & \textbf{1} & \textbf{2} & \textbf{3} & \textbf{4} & \textbf{5} & \textbf{6} & \textbf{7} & \quad \: n \quad \: \\
\hline \hline
\textbf{Traffic} ($f_{k}$) & 1317 & 239 & 42 & 14 & 4 & 4 & 1 & 1621 \\ 
\hline
\end{tabular}
\end{center} 
\caption{\textit{Traffic data-frequencies}}
\label{traffic-data}
\end{table}
For the analysis we have considered all the available 
positive counts $n_1 ,\dots n_M$ with $M^*=M$ equal to 7
which is indeed the maximum count observed.
%$S$ equal to 7 which is the max count observed ($S=T=7$). 
%%The MCMC size is 100000 with burn in equal to 5000. 
The MCMC algorithm runs for 110000 iterations 
discarding the first 10000. 
In Figure \ref{trace} the trace plots of the three main quantities: $N$, $u$ and $m_{1}$ are shown.
It is apparent that there is a strong autocorrelation which is likely yielding 
a slow mixing of the chain 
and can affect the resulting Monte Carlo error.  

\begin{figure}[!h]
\centering
%\includegraphics[height=10.7cm, width=25cm, angle=0,
%keepaspectratio]{Comparing-TAF-Wang-2012.pdf}
\includegraphics[scale=0.58]{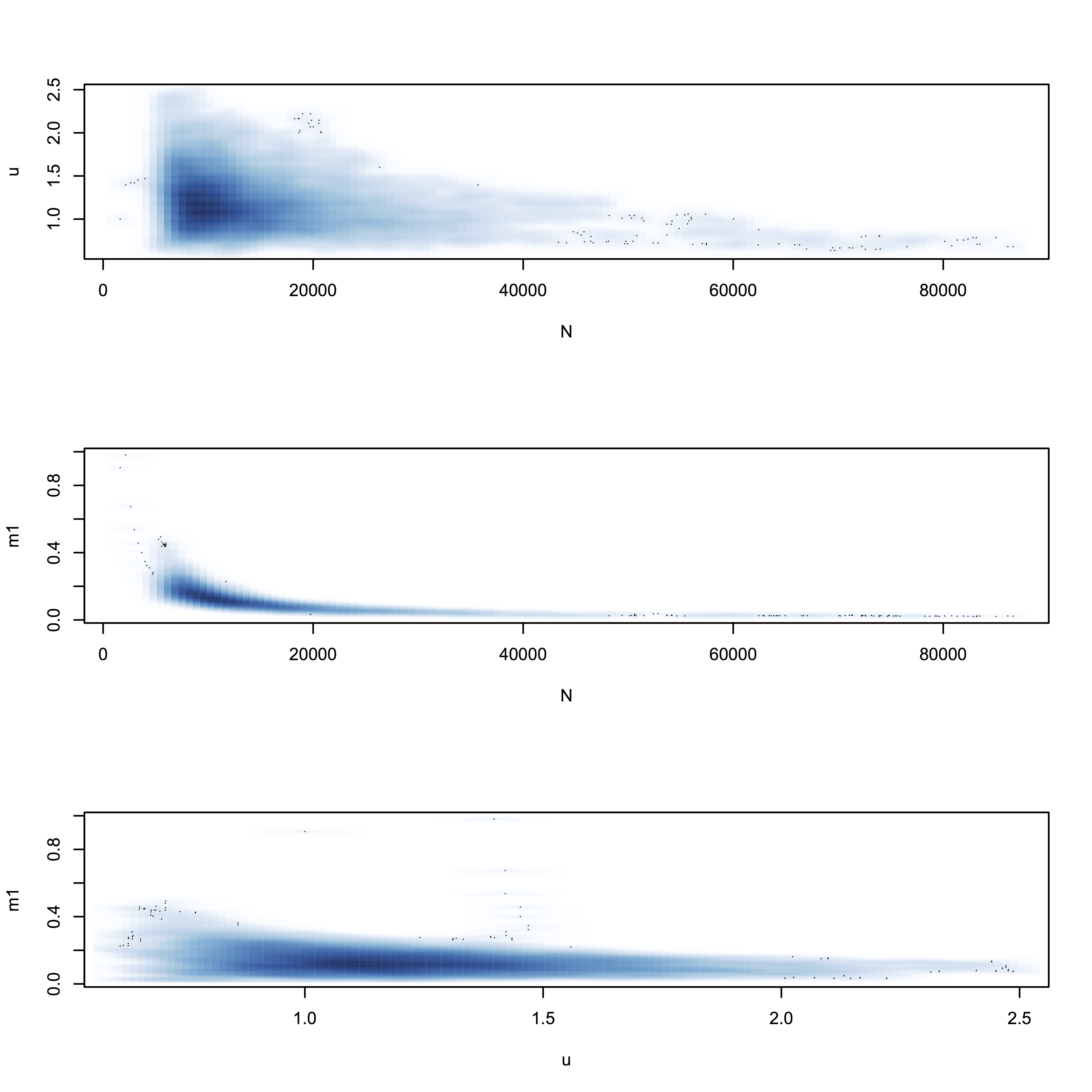}
\caption{\textit{Scatter plot of $N$, $u$ and $m_{1,7}$}.}
\label{scatterplot}
\end{figure}

\begin{figure}[!h]
\centering
%\includegraphics[height=10.7cm, width=25cm, angle=0,
%keepaspectratio]{Comparing-TAF-Wang-2012.pdf}
\includegraphics[scale=0.56]{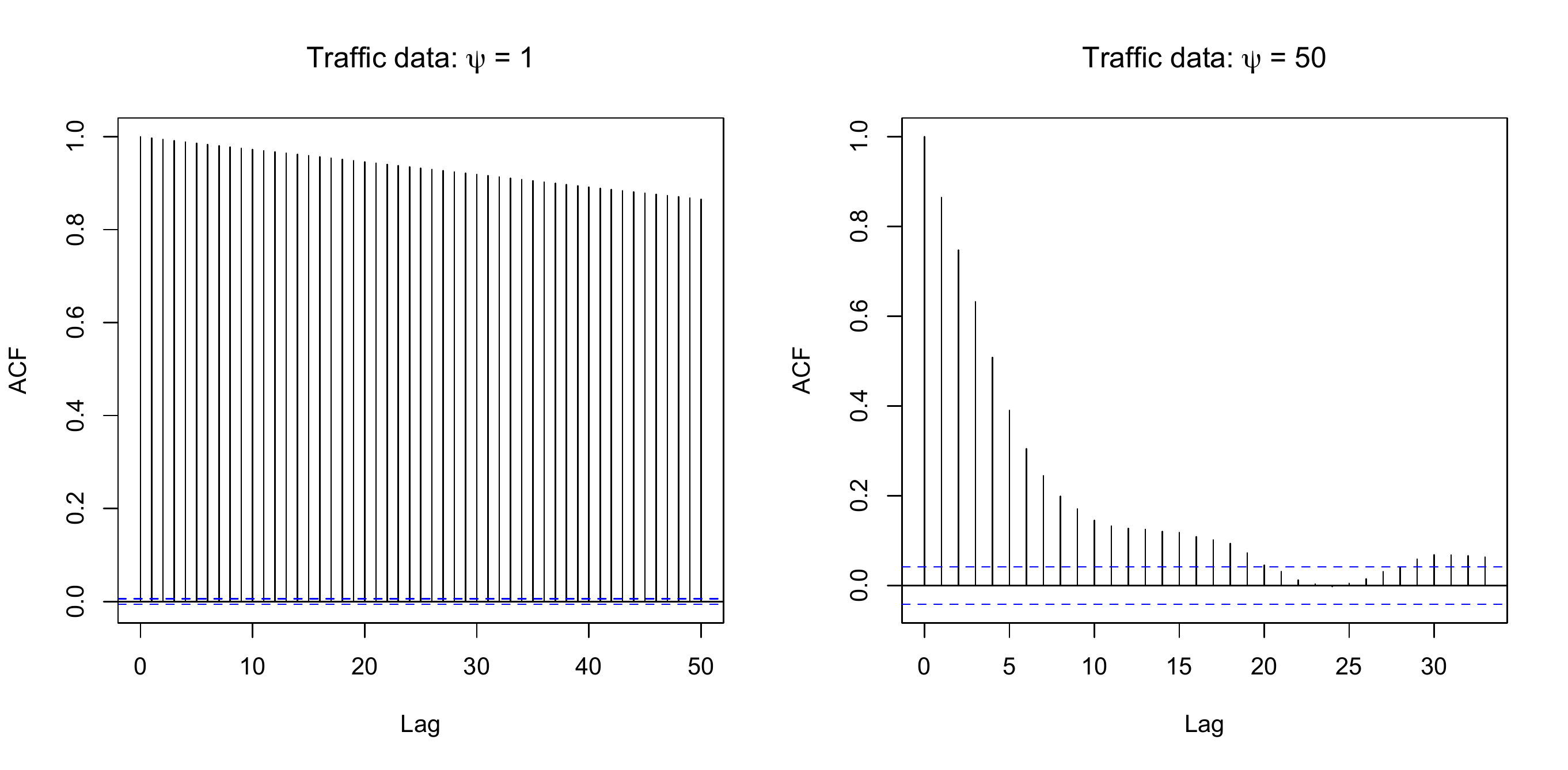}
\caption{\textit{Traffic data: $acf$ of $N$ with thin factor $\psi=1, 50$.}}
\label{acf}
\end{figure}

\begin{figure}[!h]
\centering
%\includegraphics[height=10.7cm, width=25cm, angle=0,
%keepaspectratio]{Comparing-TAF-Wang-2012.pdf}
\includegraphics[scale=0.58]{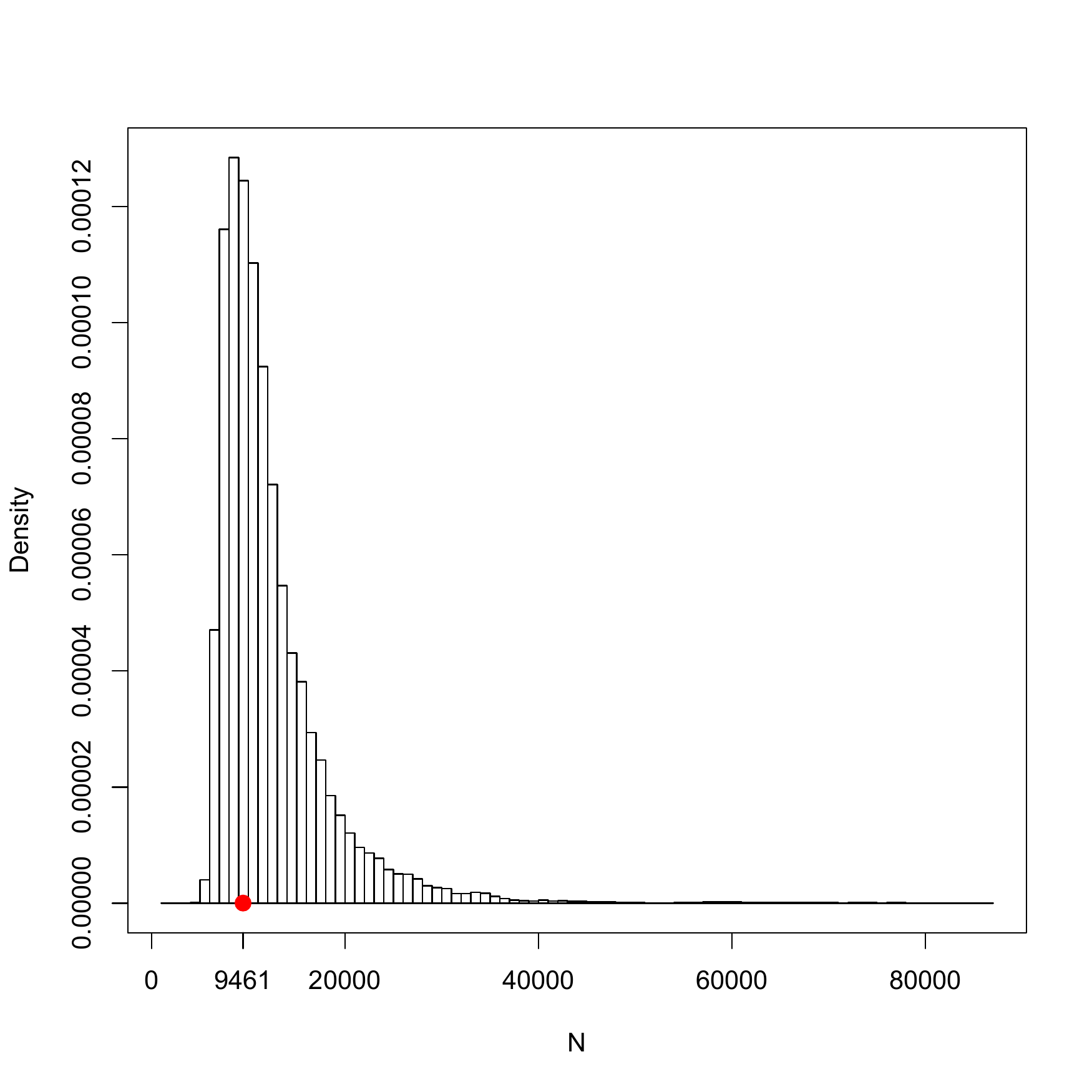}
\caption{\textit{Traffic data: Histogram of MCMC samples from the posterior distribution of $N$.}}
\label{hist}
\end{figure}

This strong autocorrelation can be due to the strong dependence among the three main quantities
as evidenced from the scatter plots in Figure \ref{scatterplot} 
(especially the one corresponding to $N$ and $m_{1}$).
However, we have verified that 
the results do not vary appreciably with a larger MCMC size. 
Indeed we redraw the acf 
considering a thin factor $\psi=50$ leading 2000 iterations.
The resulting acf in Figure \ref{acf}
looks reasonable.
As far as inference on $N$ is concerned we can see from the histogram in Figure \ref{hist} 
that the known value $N=9461$ is also very close to the mode
of the posterior distribution of $N$.
In Table \ref{traffic-results} are expressed point and interval estimates 
from different prior choices of $N$ and $u$. 
\begin{table}[!h]
\begin{center}
\begin{tabular}{|lccc|}
\hline \qquad \: \textbf{Methods} \qquad \: & \qquad \: $\hat{N}$ \qquad \: & \qquad \: $N^{-}$ \qquad \: & \qquad \: $N^{+}$ \qquad \: \\
\hline
\hline  $BPM$ & 9548 & 5642 & 22582   \\
%  $BPM_{\pi(N)\propto 1,\pi(u)\propto \frac{1}{u}}$ & 10258 & 5658 & 25859   \\ 
  $BPM_{\frac{1}{N}}$ & 9121 & 5416 & 22816   \\ 
%  $BPM_{\pi(N)\propto \frac{1}{N},\pi(u)\propto \frac{1}{u}}$ & 9433 & 5548 & 22626   \\   
  $BPM_{Rissanen}$ & 8970 & 5662 & 18255   \\ 
%  $BPM_{\pi_{Ris}(N),\pi(u)\propto \frac{1}{u}}$ & 8890 & 5692 & 18078   \\ 
 $PCG$ & 6935 & 5121 & 12843 \\
 $RBB$ & 7840 & 7742 & 7937 \\
\hline
\end{tabular}
\end{center}
\caption{\textit{Traffic data: alternative point and interval estimates}}
\label{traffic-results}
\end{table}
As we can see the point estimates are sufficiently stable 
with respect to the prior choice strategy.
Moreover our credible intervals always contain the true $N$ although the
sensitivity of the upper bound of the credible intervals seems to be
more pronounced than in the case of point estimates. \\
When we compute alternative estimators 
$\hat{N}_{PCG}$ proposed in \cite{wang:2010} and 
$\hat{N}_{RBB}$ proposed in 
\cite{rocc:bung:bohn:2011}
we have that both seem to 
be more conservative and underestimate somehow the true $N$
($6935$ and $7840$ respectively). 
However, in \cite{wang:2010} among many alternative classical procedures 
considered in that paper 
only the confidence interval derived from $\hat{N}_{PCG}$ 
through a double-bootstrap procedure gets the true $N$ inside.
Hence we consider our estimator of $N$ in this example 
one of the few successful estimators 
of the quantity of interest, in fact the closest one to the true known value.

\subsection*{Root data}
In Table \ref{root-data} are shown the \textit{Root data} already 
analyzed in \cite{wang:2010}
which represent the count distribution of the expressed genes
of the arabidopsis thaliana in the root tissue.
Notice that in this case there is a genuine interest in the unknown 
number of unexpressed genes since data are collected from a cDNA library sample
which, very likely does not allow a full screening of all expressed genes.
\begin{table}[!h]
\begin{center}
\begin{tabular}{|ccccccccccc|}
\hline
$k$  & \textbf{1} & \textbf{2} & \textbf{3} & \textbf{4} & \textbf{5} & \textbf{6} & \textbf{7} & \textbf{8} & \textbf{9} &  \\
\hline
 \textbf{Root} ($f_{k}$)& 2187 & 490 & 133 & 121 & 37 & 51 & 22 & 19 & 7 &  \\ 
%& & & & & & & & & & \\
\hline \hline
 & \textbf{10} & \textbf{11} & \textbf{12} & \textbf{13} & \textbf{14} & \textbf{15} & \textbf{16} & \textbf{17}+ & & \quad \: n \quad \:  \\
\hline
& 8 & 6 & 7 & 6 & 4 & 5 & 5 & 18 & & 3126\\
\hline
\end{tabular}
\end{center} 
\caption{\textit{Root data-frequencies}}
\normalsize
\label{root-data}
\end{table}\\
Researchers agreed that the 
arabidopsis thaliana has a relatively small genome
with approximatively 27000 protein coding genes
not necessarily all expressed in all tissues.
This information can be easily exploited in our
Bayesian procedure
formalizing an ad-hoc prior distribution for $N$ 
by setting a suitable upperbound for the population size of 
the expressed genes. We fix $N_{upp}=30000$ for our analysis.
On the other hand this (a priori) information cannot be employed so 
easily in the alternative classical approaches.
\begin{table}[!h]
\begin{center}
\begin{tabular}{|cccc|}
\hline \qquad \: \textbf{Methods} \qquad \: & \qquad \: $\hat{N}$ \qquad \: & \qquad \: $N^{-}$ \qquad \: & \qquad \: $N^{+}$ \qquad \: \\
\hline
\hline  $BPM$ & 11073 & 8739 & 15316  \\
%  $BPM_{2}$ & 11266 & 8726 & 16018 \\
 $PCG$ & 8980 & 8383 & 18771 \\
 $RBB$ & 8970 & 8652 & 9288 \\
\hline
\end{tabular}
\end{center}
\caption{\textit{Root data: alternative point and interval estimates}}
\label{root-results}
\end{table}\\
The results of the three alternative procedures are shown in Table \ref{root-results}.
As we can see the point estimates $\hat{N}_{PCG}$ and $\hat{N}_{RBB}$ are very close together
(8980 and 8870 respectively).
As argued in \cite{wang:2010} they could be a conservative 
estimate of the total number of expressed genes in the root tissue.
Our estimate is considerably higher exceeding the value 11000 for both prior choices.
Although in this case the population size is not known in advance, 
however previous works \citep{ma:2005} suggest a percentage of expressed genes in root tissue
greater than 40\% of the 
27000 protein coding genes and which fits well with the recommendation provided by $\hat{N}_{BPM}$.
  	
\subsubsection*{Colorectal polyps}
\begin{figure}[!h]
\centering
%\includegraphics[height=10.7cm, width=25cm, angle=0,
%keepaspectratio]{Comparing-TAF-Wang-2012.pdf}
\includegraphics[scale=0.98]{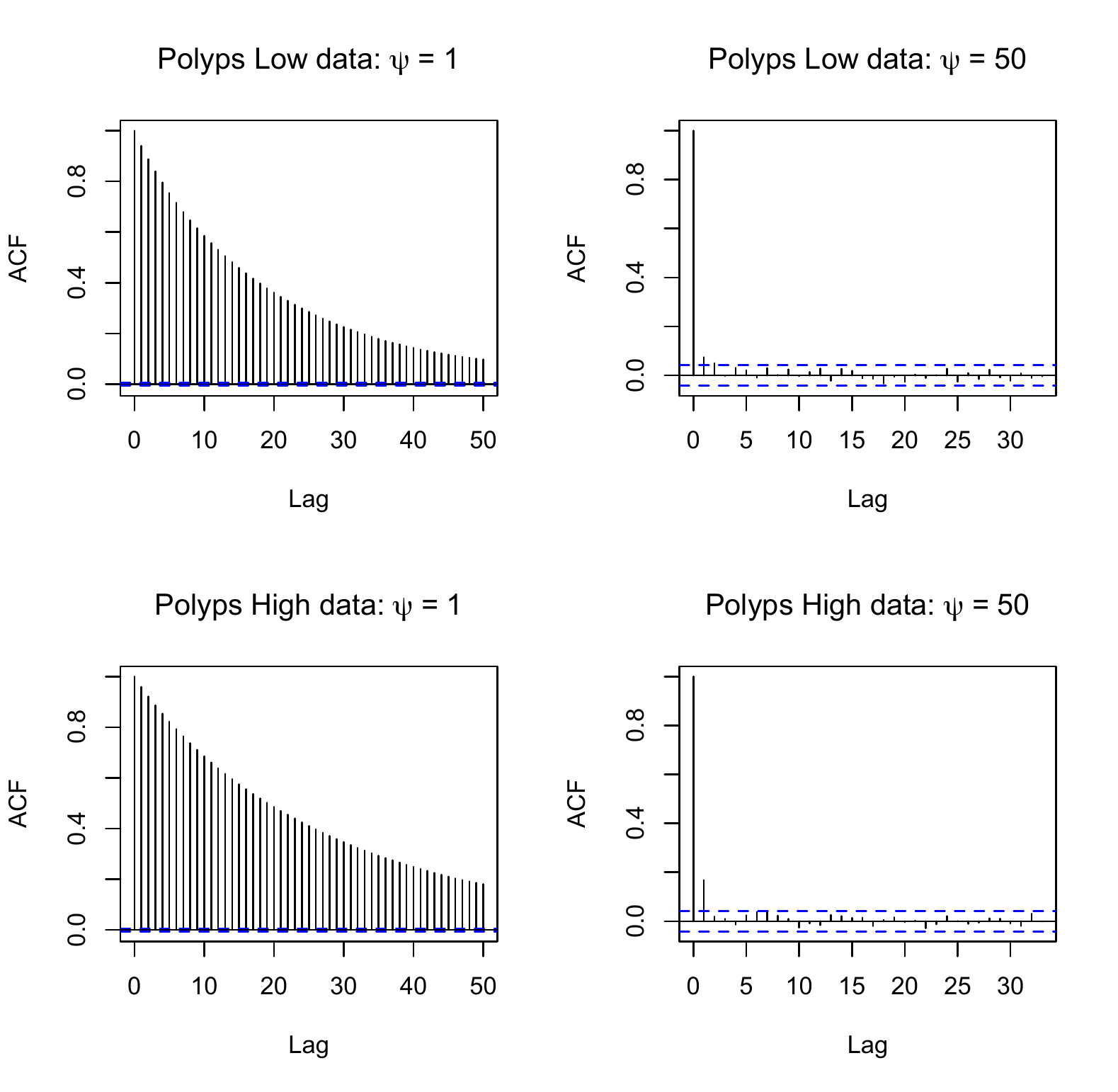}
\caption{\textit{Polyps low-high data: $acf$ of $N$ with thin factor $\psi=1, 50$.}}
\label{acf-polyps}
\end{figure}

From medical research experiences it is well recognized that diagnosing
adenomatous polyps can be subjected to undercount due to misclassification at colonoscopy.
We use data from \cite{alberts:2000} where 
in order to evaluate the recurrence of colorectal adenomatous polyps
subjects with previous history of colorectal adenomatous polyps
are allocated
to one of two treatment groups, low fiber and high fiber.
%Polyps data—frequency distribution of recurrent adenomatous polyps per patient, by
Polyps data-frequency distribution of recurrent adenomatous polyps per patient, by
treatment group is reported in Table \ref{polyps-data}.
For both groups the population size is known in advance: 584 for the
low fiber treatment ($f_{0}=285$) and 722 for high fiber treatment ($f_0 =381$) respectively.  
\begin{table}[!h]
\begin{center}
\begin{tabular}{|cccccccccccccc|}
\hline
$k$ & \textbf{1} & \textbf{2} & \textbf{3} & \textbf{4} & \textbf{5} & \textbf{6} & \textbf{7} & \textbf{8} & \textbf{9} & \textbf{10} & \textbf{11} & \textbf{12}+ & \quad \: n \quad \: \\
\hline \hline
\textbf{Polyps low} ($f_{k}$) & 145 & 66 & 39 & 17 & 8 & 8 & 7 & 3 & 1 & 0 & 2 & 3 & 299\\ 
\hline \hline
\textbf{Polyps high} ($f_{k}$) & 144 & 61 & 55 & 37 & 17 & 5 & 4 & 6 & 5 & 1 &1 & 5 & 341 \\
\hline
\end{tabular}
\end{center} 
\caption{\textit{Polyps data-frequency distribution}}
\label{polyps-data}
\end{table}\\
In Table \ref{polyps-results} are reported alternative point and the interval estimates for both 
treatments. 
In this case Wang's estimator gets closer to the true $N$
and also its confidence intervals include the main parameters of interest.
Notice that, differently from the other procedures it 
overestimate the true population size.  
%The $^{\star}$ sign means that we have used only the first 9 counts
%to compute the confidence interval values.
\begin{table}[!h]
\begin{center}
\begin{tabular}{|ccccc|}
\hline & \quad \:\: \textbf{Methods} \quad \:\: & \quad \:\: $\hat{N}$ \quad \:\: & \quad \:\: $N^{-}$ \quad \:\: & \quad \:\: $N^{+}$ \quad \:\: \\
\hline 
\hline \textbf{Polyps low} & $BPM$ & 521 & 410 & 717   \\
%  & $BPM_2$ &  &  &  \\
  & $PCG$ & 626 & 424 & 780 \\
    & $RBB$ & 492 & $446$ & $534$ \\
\hline
\hline \textbf{Polyps high} & $BPM$ & 544 & 429 & 758   \\
%  & $BPM_2$ & 546 & 429 & 764 \\
  & $PCG$ & 806 & 526 & 956 \\
  & $RBB$ & 496 & 425 & 567 \\
\hline
\end{tabular}
\end{center}
\caption{\textit{Polyps-data: alternative point and interval estimates}}
\label{polyps-results}
\end{table}\\
Our proposal, although slightly negatively biased,
yields confidence intervals which always contain the true $N$ for both data sets 
and they are also narrower than those resulting from Wang's approach.  
Moreover, as we can see from the acf plots in Figure \ref{acf-polyps} 
the autocorrelation is sensibly lower with respect to the \textit{Traffic data} example.
\subsubsection*{Scrapie in Great Britain (2002-2006)}
In Great Britain, scrapie is an endemic fatal neurological disease which affects small ruminants
(e.g. sheep, goats etc). 
In Table \ref{scrapie-data} is reported the distribution of counts of 
confirmed scrapie-affected sheep in Great Britain between 2002 and 2006 \cite{rocc:bung:bohn:2011}.
\begin{table}[!h]
\begin{center}
\begin{tabular}{|cccccccccccc|}
\hline
 $k$ & \textbf{1} & \textbf{2} & \textbf{3} & \textbf{4} & \textbf{5} & \textbf{6} & \textbf{7} & \textbf{8} & \textbf{9} & \textbf{10}+ & \quad \: n \quad \: \\
\hline
\textbf{Scrapie} ($f_{k}$) & 298 & 89 & 42 & 17 & 20 & 7 & 11 & 7 & 3 & 22 & 516 \\ 
\hline
\end{tabular}
\end{center} 
\caption{\textit{Scrapie data-frequencies}}
\label{scrapie-data}
\end{table}
For all procedures we consider the truncated distribution of the the first 9 counts
while the frequencies $f_k$ corresponding to the counts $k\geq 10$ 
are summed up to the resulting estimates. 
%%%%%%
%% QUI DICIAMO CHE AGGIUNGIAMO ALLA FINE I LEFT OUT...
%%%%%%
%adding the number of left-out (22 units) at the end.
As we can see from Table \ref{scrapie-results} the estimates produced by 
$\hat{N}_{BPM}$ and $\hat{N}_{RBB}$ are  
close together (1269 and 1220 respectively).
However, our procedure yields wider confidence interval 
compared with RBB recognizing
the possibility of more than 1500 cases of scrapie.
On the other hand, the estimates obtained by the Poisson-compound gamma approach of Wang  
appear much higher than the alternative estimators ($\hat{N}_{PCG}=1993$)
and somehow surprisingly high with respect to other recent
analyses with the same data set \citep{villas:2011}. 
Indeed, the corresponding completeness rate of 25.9\%  
seems to be too low in this case. 
Notice, however, that the point estimate returned by {\tt{pcg}} is not incompatible 
with our Bayesian inference in terms of its credible interval. 
On the other hand, the interval estimate returned by {\tt{pcg}} function
in {\tt{SPECIES}} package looks inconsistently beyond the point estimate 
possibly due to some numerical instability problems.
%Moreover, {\tt{pcg}} procedure produces an interval estimate...  
\begin{table}[!h]
\begin{center}
\begin{tabular}{|cccc|}
\hline \qquad \: \textbf{Methods} \qquad \: & \qquad \: $\hat{N}$ \qquad \: & \qquad \: $N^{-}$ \qquad \: & \qquad \: $N^{+}$ \qquad \: \\
\hline
\hline  $BPM$ & 1269 & 890 & 2165 \\
%  $BPM_{2}$ & 1272 & 893 & 2147  \\
 $PCG$ & 1993 & 4312 & 13638  \\
 $RBB$ & 1220 & 1151 & 1289 \\
\hline
\end{tabular}
\end{center}
\caption{\textit{Scrapie data: alternative point and interval estimates}}
\label{scrapie-results}
\end{table}

\subsubsection*{Methamphetamine use in Thailand}
Data in Table \ref{metha-data} is concerned with the drug abuse 
in Thailand during the last quarter of 2001. 
In this table the number of
methamphetamine users are displayed for each count of treatment episodes
reported by the public health surveillance system.  
\begin{table}[!h]
\begin{center}
\begin{tabular}{|cccccccccccc|}
\hline
$k$ & \textbf{1} & \textbf{2} & \textbf{3} & \textbf{4} & \textbf{5} & \textbf{6} & \textbf{7} & \textbf{8} & \textbf{9} & \textbf{10} & \:\: n \:\: \\
\hline \hline
\textbf{Methamphetamine} ($f_{k}$) & 3114 & 163 & 23 & 20 & 9 & 3 & 3 & 3 & 4 & 3 & 3345\\ 
\hline
\end{tabular}
\end{center} 
\caption{\textit{Methamphetamine data-frequencies}}
\label{metha-data}
\end{table}
A total of 3345 distinct drug users have been observed with maximum number of captures $M$ equal to 10. 
The count distribution has a very strongly positive skewness:
3114 out of 3345 units present only one capture.  
This is a clue for a severe undercount or, which is the same,
a large frequency $f_0$ of unreported users. 
\begin{table}[!h]
\begin{center}
\begin{tabular}{|cccc|}
\hline \qquad \: \textbf{Methods} \qquad \: & \qquad \: $\hat{N}$ \qquad \: & \qquad \: $N^{-}$ \qquad \: & \qquad \: $N^{+}$ \qquad \: \\
\hline
\hline  $BPM$ & 55435 & 35472 & 109171 \\
%  $BPM_{2}$ & 54738 & 33693 & 123273 \\
 $PCG$ & 55739 & 34783 & 93658 \\
 $RBB$ & 61133 & 60986 & 61280 \\
\hline
\end{tabular}
\end{center}
\caption{\textit{Methamphetamine data: alternative point and interval estimates}}
\label{metha-results}
\end{table}
The point estimates from Wang and B-B-R are 55739 and 61133 respectively. 
As reported in Table \ref{metha-results} our point estimate
is only slightly lower ($N_{BPM} = 55435$). 
However, similarly to Wang's procedure, our confidence interval confirms that 
there can be more than 100000 drug users.
Moreover, the lower limits of the of the interval is very close to 
Chao's lower bound 
$$
\hat{N}_{C.lb}= n+ \frac{f_{1}^{2}}{2\: f_{2}}=33090
$$
which is a conservative nonparametric estimator based on the Cauchy-Schwarz inequality.

\section{Final remarks}

We have dealt with modeling 
individual heterogeneity 
within Poisson count distribution in the absence of zero counts.
We developed an original flexible approximation of a mixture 
of Poisson distributions where 
the mixing distribution is not constrained 
to belong to a specific parametric family.\\
Our Bayesian approach described in Section \ref{sec3-moment-based} and \ref{sec3-reference-bayes}
is based on a reparameterization of the mixture likelihood function \eqref{eq1} in terms of
the first $M^*$ ordinary moment corresponding to a finite
measure $G_u$ with support $[0,u]$ where $u$ is not necessarily fixed.
In order to obtain
a probability measure 
with total mass equal to 1 we have rescaled $G_u$ to $\tilde{G}_u$ and then
we have truncated the infinite sequence of moments of $\tilde{G}_u$ to the first $M^*$
moments using an explicit renormalization which formally resembles the original likelihood \eqref{eq4}.
Moreover, we have exploited the
reparameterization of the ordinary moments into the so-called
canonical moments conveniently rescaled in $[0,1]$ allowing for an easier MCMC implementation.
Finally, in order to set-up an appropriate prior distribution on the moment space
we noted that 
conditionally on $N$ and $u$ the likelihood function has a multinomial structure
which allows us to consider a standard Jeffreys'prior 
opportunely expressed in terms of moments with the appropriate Jacobian.\\
%A best performing simulation analysis
%corresponds to a uniform prior for $N$ 
%and an invariant prior for $u$ as described in \eqref{prior-u}.  
Formal arguments and a simulation study suggested a reference 
Bayesian recipe corresponding to a uniform prior for $N$ 
and an invariant prior for $u$ as described in \eqref{prior-u}.
As shown from the simulation results our 
new fully Bayesian approach seems to perform well in terms of efficiency and coverage
although slightly more biased than Wang's estimates.
The good performances of the proposed Bayesian procedure are also 
confirmed from the results obtained in several real data analyses 
where our Bayesian approach always produced reasonable values 
for both point and interval estimates. 
Indeed for data sets where it is known in advance the population size
(Traffic and Polyps data)
the point estimates were close to the truth
and the interval estimates always contained to the true value of $N$
while for the other data-sets our proposal well agreed with 
previous scientific knowledge of the corresponding phenomenon.\\ 
The acf plots highlighted sometimes
slow convergence. 
However results obtained by our Bayesian procedure 
seem to be sufficiently stable and reliable.
Our analysis is computationally more intensive than Wang's procedure
for point estimates but lighter for interval estimates
since it relies on a costly
bootstrap procedure.\\
As future work, it would be interesting to explore 
the asymptotic behaviour 
of the procedures for $N\rightarrow\infty$. 
As argued in \cite{mao:lind:2007}, we do not have to expect 
good results from conditional likelihood approach, especially in terms of the
coverage of the interval estimates. 
However, in the examples proposed for $N$ in the range of thousands
our estimates behave reasonably well and candidates itself to 
be a good alternative to the recent $N_{PCG}$ estimator recently 
proposed by Wang.

\bibliographystyle{plainnat}
\bibliography{bpm}

\end{document}